# SKIN LAYER OF BiFeO$_3$ SINGLE CRYSTALS

Xavi Martí[1,†], Pilar Ferrer[2,3], Julia Herrero-Albillos[4], Jackeline Narvaez[5], Vaclav Holy[1], Nick Barrett[6], Marin Alexe[7], Gustau Catalan[5,8‡]

[1] Charles University in Prague, Faculty of Mathematics and Physics, Czech Republic
[2] SpLine (BM25), ESRF, Grenoble, France .
[3] Instituto de Ciencia de Materiales de Madrid ICMM-CSIC, Madrid, Spain.
[4] Helmholtz-Zentrum Berlin für Materialien und Energie GmbH, Albert-Einstein-Str. 15, 12489 Berlin, Germany
[5] Centre d'Investigacions en Nanociencia i Nanotecnologia (CIN2), CSIC-ICN, Campus Bellaterra, 08193 Barcelona, Spain
[6] CEA , IRAMIS, SPCSI, LENSIS, F-91191 Gif-sur-Yvette, France.
[7] Max Planck Institute of Microstructure Physics, Weinberg 2, 06120 Halle, Germany
[8] Institut Catala de Recerca i Estudis Avançats (ICREA), Catalunya, Spain

## **Abstract**

A surface layer ("skin") that is functionally and structurally different from the bulk was found in single crystals of BiFeO$_3$. Impedance analysis indicates that a previously reported anomaly at $T^* \sim 275 \pm 5$ $^o$C corresponds to a phase transition confined at the surface of BiFeO$_3$. X-ray photoelectron spectroscopy and X-ray diffraction as a function of both incidence angle and photon wavelength unambiguously confirm the existence of a skin with an estimated skin depth of few nanometres, elongated out-of-plane lattice parameter, and lower electron density. Temperature-dependent x-ray diffraction has revealed that the skin's out of plane lattice parameter changes abruptly at $T^*$, while the bulk preserves an unfeatured linear thermal expansion. The distinct properties of the skin are likely to dominate in large surface to volume ratios scenarios such as fine grained ceramics and thin films, and should be particularly relevant for electronic devices that rely on interfacial couplings such as exchange bias.

---

[†] xavi.mr@gmail.com
[‡] gustau.catalan@cin2.es

Bismuth ferrite, BiFeO$_3$, has become the cornerstone of magnetoelectric multiferroic research, thanks to its high ferroelectric and magnetic ordering temperatures, large ferroelectric polarization, and coupling between polar direction and magnetic easy plane, all of which are potentially useful for devices [1, 2, 3]. And, yet, the phase diagram of this archetypal material is still unresolved, with a large (and growing) number of structural and/or functional anomalies reported as a function of temperature that may or may not signal the existence of true phase transitions (see Ref. [3] for a critical review of some of them). Another unaddressed issue concerns the existence or otherwise of a surface layer (a "skin") in BiFeO$_3$. Notice that several important perovskites (e.g. SrTiO$_3$ [4,5], BaTiO$_3$ [6, 7] or relaxor Pb(Mg$_{1/3}$Nb$_{2/3}$)O$_3$ [8]) are known to have surface layers that are structurally different from the interior of the crystal. The existence of a surface layer with different symmetry in BiFeO$_3$ would not be surprising: recent work by Dieguez *et al.* [9] has shown a plethora of different structural phases within just 100 meV of the ground state of BiFeO$_3$, an energy gap comparable to surface relaxation energies of perovskites [10, 11]. As we shall show in this paper, resolving the phase diagram of BiFeO$_3$ and unravelling its surface structure are in fact linked problems: the surface itself has been found to have its own symmetry and undergo its own phase transitions which interfere with measurements of bulk BiFeO$_3$. Understanding the BiFeO$_3$ skin is also crucial from an applied point of view, because the spintronic devices proposed for this material rely on interfacial interactions, such as magnetic exchange bias, between its surface and contiguous layers [2, 12, 13].

Here we have investigated the electronic and structural properties of the skin layer of BiFeO$_3$ single crystals as a function of temperature, with techniques allowing a tuneable information depth. We show that a previously reported anomaly at T$^*$ = 275 ± 5 $^\circ$C [14] corresponds to a phase transition localized only at the interface, which explains why it was only observed by surface-sensitive probes such as back-scattering Raman [14] but not by bulk-sensitive ones such as transmission optical birefringence or neutron diffraction [15, 16]. This surface phase transition is unprecedented but may not be unique: we recall that, at low temperatures, other unexplained anomalies are also reported in this material [17, 18]. Meanwhile, at room temperature, X-ray diffraction measurements were performed both as a function of incident angle and wavelength in order to experimentally circumvent the unavoidable refraction contributions in grazing incidence geometry and reveal the true structure of the skin of BiFeO$_3$, which is found to have an elongated out-of-plane lattice parameter (tensile strain of 0.7%) in a few nanometres thick region at the topmost interface. Impedance spectroscopy and X-ray photoelectron spectroscopy show that the surface layer of BiFeO$_3$ is charge-depleted.

The crystals were grown following the method proposed by Kubel and Schmid [19]. The sample used for x-ray investigations was rosette-like with (001) habitus, as described by Burnett *et al.* [20]. The crystal was first optically polished using sandpaper and diamond paste; subsequently, in order to minimize the mechanically damaged surface layer whilst preserving its smoothness, a 30 minute chemical-mechanical polishing using silica slurry (Syton SF1; Logitech) was employed. The pressure applied during the last stage of the mechano-chemical polishing was less than 100 grams per cm$^2$ so as to minimize polishing-induced stress at the surface. The final local root mean square roughness was 0.8 nm, as determined by atomic force microscopy, and in agreement with X-ray reflectivity (Cu k$_\alpha$) fitted value of (1.3 ± 0.1) nm (Fig. 1b). The abrupt decrease of reflected intensity when surpassing the critical angle demonstrates that the sample surface is suitable for grazing incidence diffraction (GID).

Ferroelectric/ferroelastic domains are optically observable using standard birrefringence (Fig. 1a). The quadrant arrangement is consistent with the polar vector pointing alternately along the four diagonals of the perovskite unit cell, in a crystal with surface perpendicular to the $(001)_{pseudocubic}$ direction. However, the existence of ferroelectric flux closure cannot be verified by birrefringence alone, and further experiments using e.g. piezorresponse force microscopy are required to confirm this. Precedents of ferroelectric flux closure have so far been limited to nanoscopic domains in thin films [21].

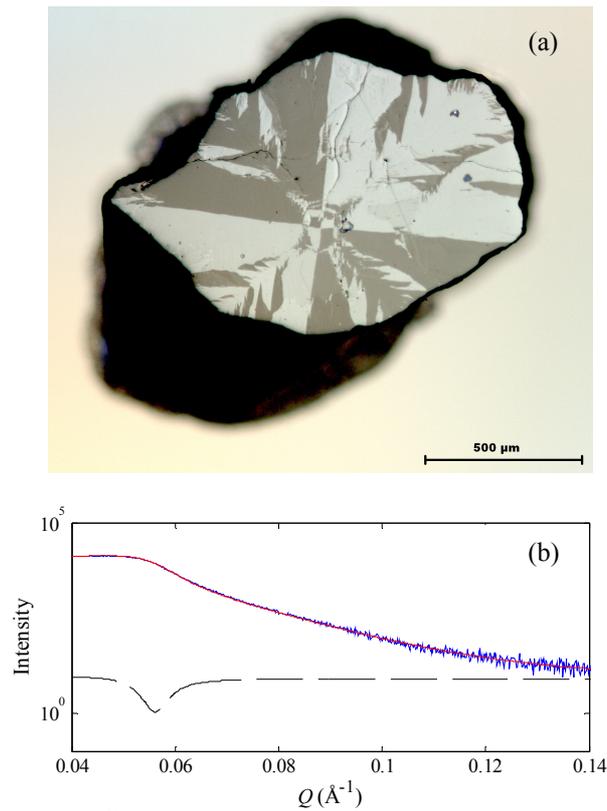

**FIG 1.** (a) Polarized optical microscopy viewgraph of a mirror-polished single crystal of BFO with habitus parallel to $<001>_{pseudocubic}$, showing a quadrant-like arrangement of the four types of ferroelastic domain present in the sample. (b) X-ray reflectivity measurement (blue) and theoretical fit assuming a roughness of 1nm. The anomaly in the first derivative of the experimental curve signals the value of the critical angle (dashed line) in agreement with reported bulk values.

Dendritic crystals with (110) planes parallel to the surface were also made as described in refs. [22-25], and their electrical properties were characterised, both for a polished sample and for an unpolished one. The ac impedance was measured using an Agilent impedance analyzer (model 4294A), and is shown in Fig. 2 (panels a and b). The electrodes were sputtered gold, with Pt wires attached to them using silver paste. The capacitive impedance shows classic Maxwell-Wagner behaviour [26]: a giant and frequency-dependent step-like increase in capacitance, typical of a material with two lossy dielectric components in series, attributed to the bulk and the interfacial regions. The large losses indicate that the crystal becomes very conducting as temperature increases, so that the voltage is then mostly dropped on a charge depleted interfacial barrier layer (contact capacitance); because the interface is thin, the *apparent* capacitance becomes very large [27, 28].

Further evidence for an electronically different surface layer comes from X-ray photoelectron spectroscopy (XPS). The Bi 4f core level results are also shown in figure 2. Spectra were acquired at normal emission and at an angle of 30° with respect to the sample surface using small spot Al Kα radiation. Two components are observed in the Bi 4f spectrum indicating two distinct electronic environments for $Bi^{3+}$. The higher binding energy component is more intense at glancing angle detection showing that it should be associated with the surface or near-surface region. No carbonate species were observed in the C 1s spectra (not shown), thus the high binding energy component is not due to adsorbates but is intrinsic to the surface. Higher binding energy is consistent with lower screening and therefore with lower electron density, as also observed in the skin of other ferroelectrics [6]. The Fe 2p spectrum was also checked and showed a spin orbit splitting (13.45 eV) and $2p_{3/2}$ binding energy (710.8 eV), fully consistent with trivalent Fe. The XPS evidence thus points to a distinct electronic environment whilst maintaining the same chemistry as the bulk BFO. Finally, no metallic Bi peak was observed showing that no significant reduction of the BFO takes place. Thus, although an electronically different surface layer is observed, the stoichiometry and the cation oxidation states do not change.

A key observation here is that, for the low frequency data (where interfacial capacitance dominates), there is a small peak whose temperature ($T^* = 272\ ^oC$) is frequency-independent (Fig. 2, panels a and b). The fact that the position of $T^*$ is frequency-independent indicates that this is not a dynamic effect but a true phase transition. On the other hand, the fact that there is no peak in the high frequency curves (where bulk capacitance dominates) implies that the bulk of the crystal is not undergoing any change [29]. The $T^*$ peak must therefore correspond to a phase transition confined within the skin of the crystal. The same anomaly is observed also in the as-grown unpolished samples (panel a, inset) and therefore this surface transition is not caused by polishing.

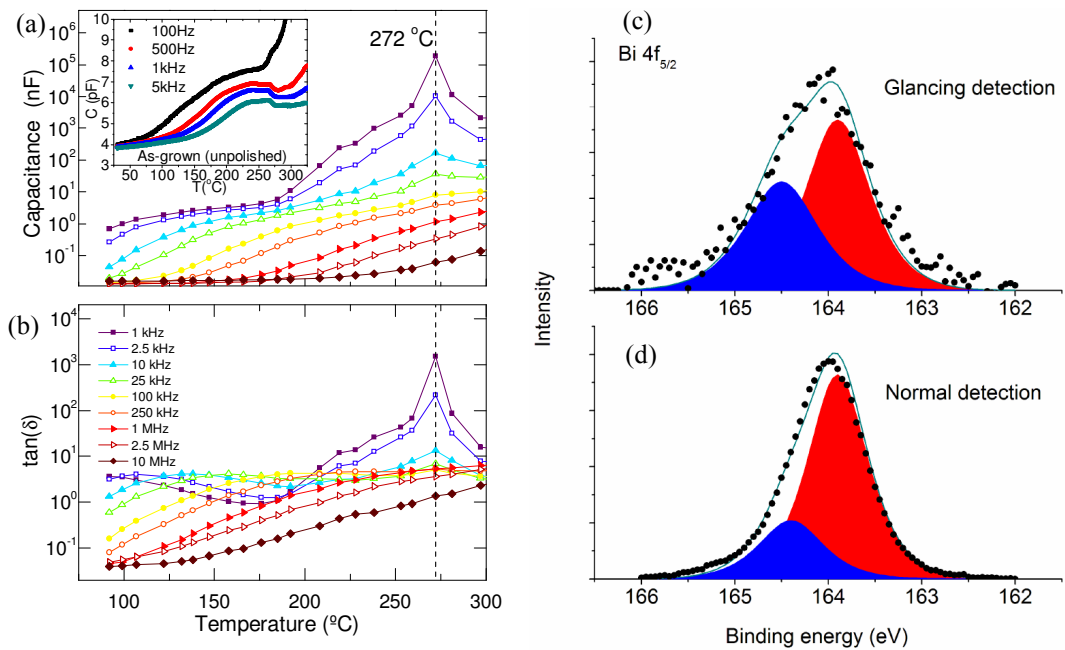

**FIG. 2.** The impedance spectroscopy (panels a and b) display a peak at T*=272 °C in the low-frequency, region of the spectrum, which corresponds to the contact capacitance. This peak, is absent at higher frequencies, so the phase transition only affects the interface and not the bulk. Inset: the anomaly was also observed in unpolished crystals, so the surface distortion is not caused by the polishing process. Panels c and d: the XPS intensity at glancing detection (c) is shifted towards higher energies than that at normal incidence (d), consistent with a charge-depleted surface.

On these grounds, grazing incidence X-ray diffraction (GID) is a very convenient technique to track the structural changes at the surface, due to its tuneable information depth from few unit cells to several hundreds of microns by changing either the incidence angle or the wavelength [30]. The same $BiFeO_3$ crystal shown in Fig. 1a was investigated by GID on the six-circle diffractometer at SpLine beamline, ESRF, Grenoble, France [31]. The sample was placed on a heating stage (25 °C - 400 °C) covered with an air-tight capton housing filled with 1 bar of pure oxygen to minimize oxygen vacancies.

In order to tune the penetration depth we have devised a dual approach, changing both the incidence angle and the incidence photon energy. The reason for this stems from the fact that, while either parameter alone would permit probing different depths (see figure 3-b), they would also yield refraction-induced peak shifts [32], so that both the real (structural) shifts and the artificial (refractive) ones are convoluted. The key to deconvolute the refraction corrections is to note that these behave very differently as a function of angle and energy (see figure 3c), especially at the vicinity of the critical angle and around a absorption edge (Bi $L_3$ at ~13 keV). In particular, inspection of Fig. 3b and 3c shows that for an incidence angle of 0.22 °, the L shift due to refraction is exactly the same for 10 keV and 15 keV, both achievable at the Spline beamline [31]. Although this set of angles and energies is not unique, it is convenient because it will allow discriminating penetration depths between few Angstroms and several tens of nanometres.

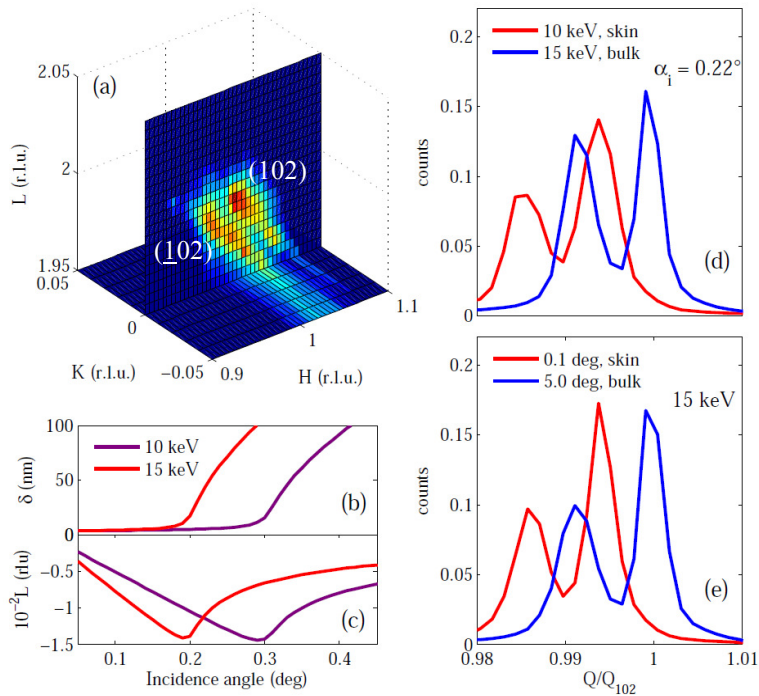

**FIG. 3:** (a) Reciprocal space map around the (102) reflection. The penetration depth (b) and refractive shift (c) as a function of incidence angle for two different beam energies. The $Q$-histogram analysis of the HKL maps measured by changing the photon energy (d) and the incidence angle (e)

Reciprocal space maps were collected around the (102) reflection. We used a pseudo-cubic unit cell to explore the reciprocal space using a base of three perpendicular reciprocal lattice vectors: H, K and L, being the latter the direction perpendicular to the crystal surface. Three dimensional meshes were collected to capture all diffracted intensity around the chosen (102); in figure 3a we show two perpendicular sections of this three-dimensional mesh. We remark the following two features: first, due to the ferroelastic twinning, two reflections, (102) and (0$\underline{1}$2), are simultaneously observed. Second, there is considerable broadening of the intensity maxima due to mosaicity [33], making it difficult to track an individual diffraction maximum. However, the mosaic blocks are randomly rotated but not deformed, i.e., the actual lattice parameters are independent of mosaic rotation (the mosaic broadening of the diffraction maxima is in the direction perpendicular to the scattering vector). We therefore propose to track the modulus of the momentum transfer $Q = \sqrt{H^2 + K^2 + L^2}$, in analogy with powder diffraction, and use its changes to identify the structural distortions.

In Figs. 3d and 3e we plot the radial distribution of intensities integrated for each shell of constant $Q$ value (i.e., the histogram of the moduli of $Q$) for the 3-dimensional HKL maps. We summarize here the experimental conditions: panel d shows data acquired at same incidence angle (0.22 °) but different photon energies (10 and 15 keV); in panel e the photon energy is constant (15 keV) and the incidence angle is varied (0.1 ° and 5 °). For each panel, red curves denotes the surface of the crystal while the bulk contribution is represented by the blue curves. We normalized for the total number of counts to allow a comparison between red and blue curves. We shifted the histograms according to refraction [32]. Data show an unambiguous difference between surface and bulk. Notice that such difference cannot be explained by refraction as the shift of the histograms was exactly the same for all data in panel (d) and, moreover, the applied corrections in panel (e) lead to the very same results. Therefore, there is an expansion of the (102) interplanar distance located at the topmost few nanometres of the crystal. Because the skin must be in-plane coherent with the bulk, this result indicates that the out-of-plane lattice parameter of BFO is elongated by as much as 0.7% at the surface.

The temperature dependence of the out-of-plane lattice parameter was also measured and is shown in Fig. 4 for the bulk and for the surface (incidence angle of 0.1 degree, penetration depth of 1nm). The difference between the two curves is striking: whereas the bulk shows a fairly featureless linear thermal expansion, the curve corresponding to the skin shows a marked anomaly, with negative thermal expansion setting in around 260°C, followed by a sharp expansion of almost 0.2% at 280°C. The critical temperature is essentially the same as measured by impedance analysis. Its detection only at sub-critical grazing incidence confirms that this phase transition is confined within a skin layer which is less than 10nm thick.

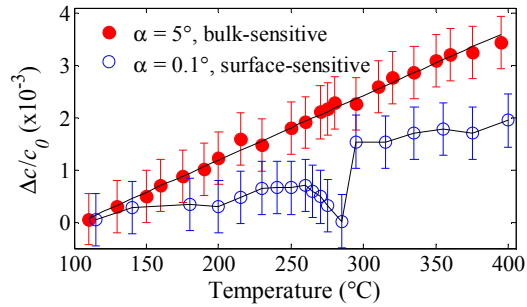

**FIG. 4.** Comparison between the reciprocal thermal expansion at the skin (solid symbols) and inside the crystal (empty symbols) evidencing the local phase transition at T* in the surface of $BiFeO_3$.

This surface-confined transition is facilitated by the existence of many metastable phases within 100meV of the groung state of $BiFeO_3$ [9]. The energy barrier between these phases may be easily overcome by surface tension and the observed charge depletion. We parenthetically note that the (100) planes of $BiFeO_3$ are not charge neutral, so minimization of the electrostatic cost of the surface does require some form of surface reconstruction and/or electronic reconfiguration. We do not discard that skin effects also participate in other unexplained phase transitions of this material[3]: two prime candidates are those at 200K and 140K, which, like $T^*$, have also been detected by surface-sensitive probes such as back-scattering Raman [17, 18] but not by bulk transmission neutron diffraction or magnetometry [35].

The structural and functional behaviour of the skin is likely to affect the performance of interfacial coupling devices such as those based on exchange bias. It will also affect the phase diagram of very thin films: the results suggest that, as well as strain[34], finite size effects should be included in their analysis -we note that indeed the $T^*$ anomaly has been reported thin films [36]. Given the importance of its repercusions, we believe the surface of $BiFeO_3$ deserves much closer scrutiny from the multiferroic community.

We thank Professor Hans Schmid for the loan of one of the crystals used in this study, and for the critical reading of this manuscript. We also thank the SpLine staff for their assistance in using beamline BM25B-SpLine and Pascale Jégou for the XPS measurements. We acknowledge the financial support of the Spanish Ministerio de Ciencia e Innovación (PI201060E013), Consejo Superior de Investigaciones Científicas (PIE 200960I187), the German Science Foundation (Grant SFB762), Ministry of Education of Czech Republic (MSM0021620834), the Grant Agency of the Czech Republic (P204/11/P339), and of the EU (Project NAMASTE, 214499). M. A. and G. C. thank the Lewerhulme trust for the funds that have enabled their collaboration.